\documentclass[aps,prl,twocolumn,superscriptaddress,floatfix,footinbib,10pt]{revtex4-2}

\usepackage[utf8]{inputenc}

\usepackage{graphicx}
\usepackage{float}
\usepackage{bm}
\usepackage{amsmath}
\usepackage{wasysym}
\usepackage{amsfonts}
\usepackage{mathrsfs}
\usepackage{color}
\usepackage[colorlinks,linkcolor=blue,urlcolor=blue,citecolor=blue]{hyperref}

\begin{document}

\title{Thermal Quench Dynamics of Visons in Gapless Kitaev Spin Liquid}

\author{Yang Yang}
\affiliation{Department of Physics, University of Virginia, Charlottesville, Virginia, 22904, USA}

\author{Gia-Wei Chern}
\affiliation{Department of Physics, University of Virginia, Charlottesville, Virginia, 22904, USA}

\begin{abstract}
The relaxation dynamics of the Kitaev honeycomb model under a thermal quench is dominated by the quasi-stochastic diffusion and pair annihilation of visions, which are gapped flux excitations of an emergent $\mathbb{Z}_2$ gauge field of the Kitaev spin liquid. Both the diffusion energy barrier as well as the effective interactions between visons are mediated by the Majorana fermions which are fractionalized quasiparticles of the spin liquid. Through extensive kinetic Monte Carlo simulations, we show that the interplay between the thermal diffusion and nonlocal multi-vision interactions leads to a variety of temperature-dependent dynamical behaviors ranging from diffusion-limited and terminal-velocity-limited annihilation to dynamical arresting and freezing.  Notably, we show that the freezing phenomenon is intimately related to the formation of metastable $\sqrt{3}\times\sqrt{3}$ vison crystals and a hidden coarsening of super-clusters associated with a broken $\mathbb{Z}_3$ symmetry. 
\end{abstract}
\date{\today}

\maketitle

{\it Introduction} -- The Kitaev honeycomb model is one of the primary playgrounds for studying quantum spin liquids with fractionalized excitations~\cite{Kitaev2006}. The great interest in this system is partly due to its exact solvability~\cite{Nasu2014,Nasu2015,Feng2020,Sengupta2008,Patel2012,Bhattacharya2016,Sameti2019,Nasu2019,Rademaker2019,Zhu2021,Jin2023,Roberts2024,Motome2020} and potential material realizations~\cite{Jackeli09,Takagi19,Trebst2022}. The excitations of the spin-liquid ground state are gapless Majorana fermions couple to an emergent $\mathbb{Z}_2$ gauge field~\cite{Kitaev2006,Lieb1994}, of which the elementary excitations are visons that carry $\pi$ fluxes of the gauge field. The fractionalization also manifests in the thermodynamic behaviors of the Kitaev system~\cite{Nasu2014,Nasu2015,Feng2020}. The $\ln2$ entropy density of Ising spins are released in two stages when the system is cooled from the paramagnetic phase. The freezing of Majorana fermions at the temperature scale of exchange interaction leads to the release of exactly half the total entropy. The remaining half of the entropy associated with the $\mathbb{Z}_2$ fluxes is released at a lower temperature set by the energy scale of vison interactions. The Kitaev model is thus one of the few quantum spin liquids that a full theoretical modeling of its thermodynamic behaviors is possible. However,  the fundamental thermal relaxation dynamics of the Kitaev spin liquids remains an open question.

The relaxation or phase ordering dynamics of systems with broken symmetries is an important subject in nonequilibrium physics with a long history \cite{Hohenberg1977,Bray1994}. Several universality classes and dynamical scaling regimes have been classified. The relaxation towards long-range or quasi-long-range ordered ground state is often dominated by the nonequilibrium dynamics of topological defects of the order-parameter field.  For instance, the phase ordering of an XY model subjected to a thermal quench is governed by the annihilation dynamics of vortices~\cite{Jelic2011}. In contrast, for systems without symmetry breaking, such as classical or quantum spin liquids, the characterization of their relaxation dynamics is more subtle. For an important class of such systems which exhibit fractionalization phenomena, their phase ordering dynamics is similarly controlled by the annihilation process of the fractionalized quasiparticles. For example, the growth of correlation length in a classical spin ice is shown to be controlled by the annihilation of magnetic monopoles, which are emergent particle-like objects that violate the ice-rule constraints~\cite{Castelnovo2010}. 

    \begin{figure}[ht]
        \includegraphics[width=\linewidth]{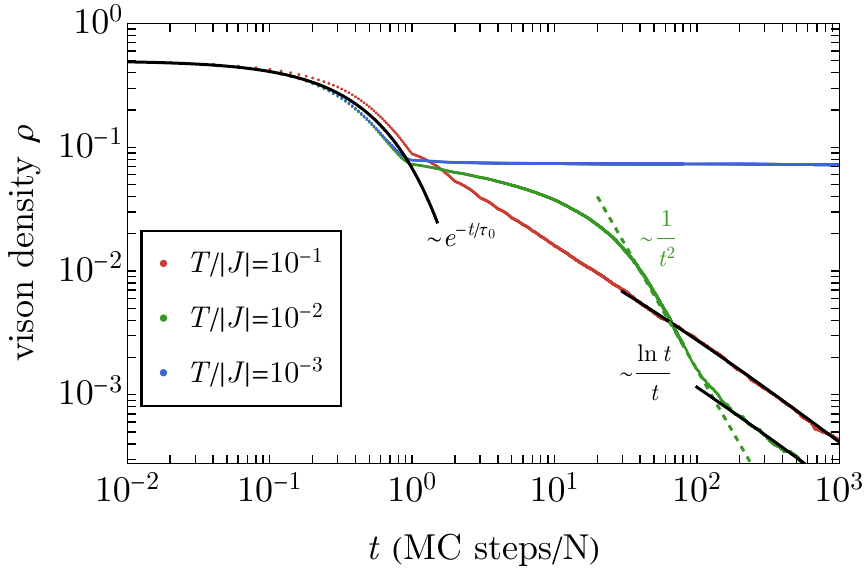}
        \caption{\label{fig:1} Time evolution of the vison density $\rho$ after the thermal quench at various post-quench temperatures: $T/|J|=10^{-1}$ (red), $T/|J|=10^{-2}$ (green), $T/|J|=10^{-3}$ (blue) obtained from $N=60\times 60$ systems averaged over $100$ runs for each temperature.}
    \end{figure}

In this work, we present a comprehensive theory for the phase ordering of  Kitaev spin liquid in the isotropic limit, focusing on the annihilation dynamics of visons. 
Our main results, summarized in Fig.~\ref{fig:1} which shows the time evolution of the vison density $\rho(t)$ after a thermal quench, reveal three distinct dynamical behaviors of visons depending on the post-quench temperatures. At high temperatures, the asymptotic $\rho \sim t^{-1} \ln t$ decay indicates a diffusion-limited annihilation of nearly free visons. Conversely, at very low temperatures, the vison density quickly freezes to a constant after a fast initial decay. The freezing behavior is shown to be intimately related to the formation of metastable vison crystals. Finally, an accelerated annihilation due to an attractive vison interaction gives rise to a power-law $\rho \sim t^{-2}$ decay at intermediate temperatures. 
In the following, we elucidate the origins of these three distinct types of vison dynamics and the corresponding temperature scales.

{\it Simulation method} -- The exact solvability of the Kitaev model is intimately related to its static $\mathbb{Z}_2$ gauge field. As a result, similar to classical Ising spins, there is no fundamental dynamics for visons in the Kitaev spin liquid. 
However, effective dynamics of visons can be induced by perturbations that break the integrability of the Kitaev model~\cite{Joy2022,Chen2023}. In the thermal quench scenarios considered here, such perturbations can come from couplings to a heat bath. Formally, the dynamical evolution of such an open quantum system from time $t_1$ to $t_2$ can be described by quantum operations or quantum dynamical map formalism~\cite{Sudarshan61,Nielsen_Chuang_2010}: $\rho(t_2) = \mathcal{E}[ \rho(t_1) ]$, where $\rho(t)$ is the density matrix of the open quantum system under consideration, which is the Kitaev model in our case.

The coupling to the heat bath introduces additional energy or time scales to the problem. Phenomenologically, there are three reservoir-induced time scales: $\tau_{\rm M}$ characterizes the relaxation time of the Majorana fermions, $\tau_2$ denotes the decoherence time of visons, and $\tau_1$ corresponds to the inverse hopping rates of visons. The gapless and continuous spectrum of Majorna fermions implies a more efficient coupling to the heat bath and a faster relaxation time compared with the time scale of vison dynamics, i.e. $\tau_{\rm M}\ll \tau_{1}, \tau_2$. We thus assume quasi-equilibrium Majorana fermions with respect to instantaneous vison configurations in dynamical simulations of visons, an approach similar to the Born-Oppenheimer approximation in quantum molecular dynamics~\cite{marx09}. We further assume a fast decoherence time $\tau_2 \ll \tau_1$, which means superposition of vison states is quickly destroyed by thermal couplings. 
The density matrix of the Kitaev system during the relaxation is then described by a product form: $\rho = \rho_{\rm V} \otimes \rho^{\rm th}_{\rm M}$, where $\rho_{\rm V}(t) = \sum_{\mathcal{C}} \pi\left(\mathcal{C}; t \right) |\mathcal{C} \rangle \langle \mathcal{C} |$ is the reduced density matrix which is diagonal in the vison configuration basis $|\mathcal{C}\rangle$, and $\rho^{\rm eq}_{\rm M} = \exp\left[-\beta \hat{\mathcal{H}}_{\rm M}(\mathcal{C}) \right]/\mathcal{Z}_{\rm M}$ is the equilibrium Majorana density matrix, and $\mathcal{H}_{\rm M}(\mathcal{C})$ is the quadratic Majorana Hamiltonian which depends on the instantaneous vison configuration. Here a vison configuration is specified by $\mathcal{C} \equiv \{W_\alpha\}$, with $W_\alpha = \pm 1$ denoting the flux of honeycomb plaquette~$\alpha$.

Focusing on the vison sector,  the diagonal matrix element $\pi(\mathcal{C}; t)$ denotes the time-varying probability density of a given vison configuration. A discrete-time dynamical evolution of visons can be modeled by a set of Kraus operators $\rho_V(t + \delta t) = \sum_\eta K_\eta \rho_{\rm V}(t) K^\dagger_{\eta}$~\cite{Kraus1983}, where the time-step $\delta t$ is of the order of $\tau_1$, and each $K_\eta$ provides one channel of transition between two vison configurations differed only by single vison hopping such that application of Kraus operators preserves the diagonal form of $\rho_{\rm V}$. 
This discrete dynamics can also be effectively formulated as a Markov chain process for the vison configuration probability: $\pi(\mathcal{C}, t+\delta t) = \sum_{\mathcal{C}'} P(\mathcal{C}' \to \mathcal{C}) \pi(\mathcal{C}'; t)$. The specific form of the transition matrices $P$ depends on microscopic details of the heat-bath coupling. As the system is driven towards thermal equilibrium by the heat bath, we employ a Glauber-type transition probability $P(\mathcal{C}' \to \mathcal{C}) \propto 1/[1 + \exp(\Delta E/ T)]$, which depends on the energy difference $\Delta E = E(\mathcal{C}') - E(\mathcal{C})$ of vison configurations before and after a nearest-neighbor hop; here the vison energy is given by $E(\mathcal{C}) = \langle \mathcal{H}_{\rm M}(\mathcal{C}) \rangle = {\rm Tr}\left[ \rho^{\rm eq}_{\rm M} \mathcal{H}_{\rm M}(\mathcal{C}) \right]$. The exact diagonalization of the Majorana fermions is used to compute the expectation values, and kinetic Monte Carlo (kMC) method is adopted to simulate the Markov chain process. More details of the implementation can be found in supplemental materials (SM)~\footnote{\label{SM} Details are provided in Supplemental Materials.}.




    
We performed kMC simulations with quench temperatures across three orders of magnitude, from $T/|J|=10^{-3}$ to $10^{-1}$. The system is prepared in a state with randomly distributed visons of an initial density $\rho_0 \approx 0.5$, corresponding to initial states at infinitely high temperature. At every step of the simulation, a stochastic move to one of the six neighbors, determined by Boltzmann factors of the corresponding energy change, is applied to a randomly selected vison in the system. Pair annihilation occurs when the selected neighbor is occupied by another vison; both are then removed from the system. In order to compare results from different system sizes, the time-step $\delta t$ of a single vison update is defined as $\delta t = \tau_1/N$, where $N$ is the number of spins~\cite{stoll73,binder74}. As a result, the time-scale of a system sweep is of the order $\tau_1$, which also serves as the time units in the kMC results (with up to $N = 120^2$ spins) presented below. 


The early stage of the relaxation is dominated by the pair-annihilation of neighboring visons in the initial state, since the corresponding energy gain is approximately $0.26|J|$~\cite{Kitaev2006}, which is higher than all the post-quench temperatures considered here~\footnote{Above $T/|J|=0.26$, it is expected that the system will exhibit pair-production of visons, giving rise to the high-temperature peak in the specific heat of the Kitaev model \cite{Nasu2014}. As we focus on temperatures well below this threshold, we exclude the pair-production of visons in our simulations to concentrate solely on the annihilation dynamics of visons.}. The annihilation of vison pairs can be described by a rate equation $d\rho_{\rm pair} / dt = -\rho_{\rm pair} / \tau_0$, where $\rho_{\rm pair}$ is the density of vison pairs and $\tau_0 \approx 2.0\pm 0.2$ is an effective lifetime. This accounts for the initial exponential decay of vison density observed for all three temperatures shown in Fig.~\ref{fig:1}.

{\it Diffusion limited annihilation} -- At quench temperatures higher than the energy scale of vison-vison interactions $T_{\rm int} \sim 0.04|J|$, the random walks of visons are essentially unbiased with equal probabilities towards any of the six neighbors. The temperature scale $T_{\rm int}$ is estimated from the difference between the energy cost of creating two well-separated visons ($0.3|J|$) and the energy gain of pair annihilation from two neighboring visons ($0.26|J|$)~\cite{Kitaev2006}. The relaxation process in this regime is precisely the same as that of the single-species diffusion-limited reaction system $A+A\rightarrow\emptyset$ in 2D, which has been widely studied~\cite{Toussaint1983,Amar1990,Lindenberg1995,Ginzburg1997,Sherrington2002,Krapivsky2010}.  A coarse-grained theory for the spatiotemporal evolution of visons is given by the reaction-diffusion equation~\cite{Krapivsky2010}
    \begin{align}\label{eqn:reaction-diffusion}
        \frac{d\rho}{dt}= D\nabla^2 \rho- K\rho^2,
    \end{align}
where $D$ is the diffusion coefficient and $K$ is the reaction rate. In the homogeneous limit, Eq.~(\ref{eqn:reaction-diffusion}) yields $\rho\sim t^{-1}$ for $d=2$. However, since $d=2$ is the critical dimension for the mean-field treatment, the power-law behavior is modified by a logarithmic correction~\cite{Yurke1993,Shimizu2023,Krapivsky2012}. Our simulation results at $T \sim 10^{-1}|J|$ agree very well with the $\rho \sim t^{-1}\ln t$ asymptotic behavior of the vison density, confirming the unbiased random walks of visons and diffusion-limited annihilation dynamics in the high temperature regime~$T_{\rm pair} \gtrsim T \gtrsim T_{\rm int}$, where $T_{\rm pair} \approx 0.26$ corresponds to the energy scale of pair production.

\begin{figure}[t]
        \includegraphics[width=\linewidth]{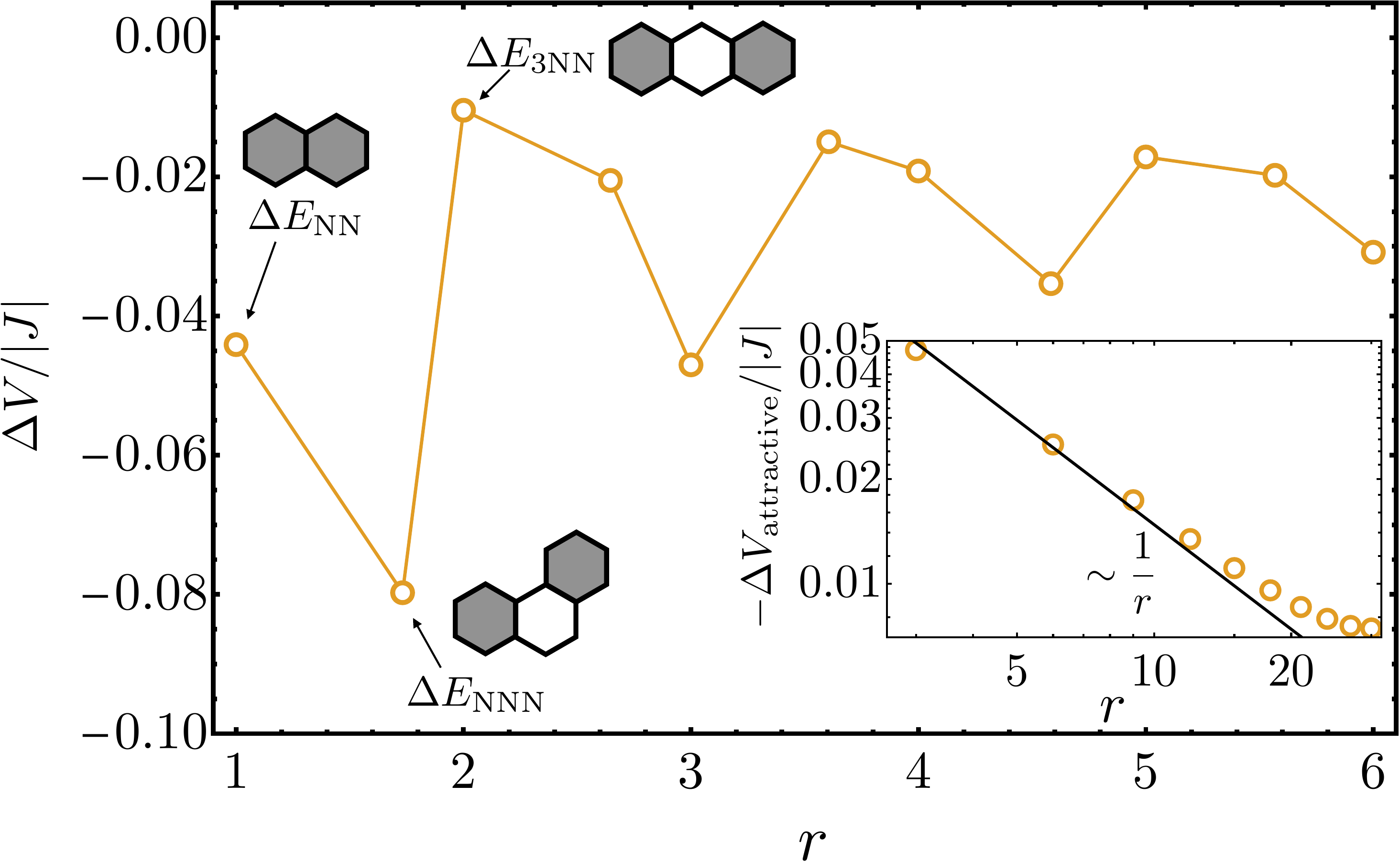}
        \caption{Effective interaction potential between two visons as a function of their separation $r$ obtained from an $N=60\times 60$ system. The minimum interaction potential occurs at $r=\sqrt{3}$ for the next-nearest-neighbor (NNN) pair. The inset shows the overall attractive interaction for two visons features $1/r$ dependence before the finite-size effect comes to play at large $r$ in the log-log scale \cite{Note1}. \label{fig:Vr}}
    \end{figure}

{\it Terminal-velocity limited annihilation} -- Upon lowering the temperature below $T_{\rm int}$, the dynamics of visons becomes increasingly affected by their interactions, which exhibit a complex dependence on vison configurations. Notably, the interaction potential between a pair of visons, as shown in Fig.~\ref{fig:Vr}, features an overall attractive $1/r$ dependence on the distance with a short-period oscillatory behavior~\cite{Kitaev2006}.  As will discussed in the following, the energy barriers associated with the short-distance oscillation would lead to a dynamical freezing of visons. For the intermediate temperature regime $T_{\rm int} \gtrsim T \gtrsim T_{\rm freezing}$, the vison dynamics is dominated by the long-range attractive interaction. As a result, instead of unbiased random walks, the semi-stochastic motion of a vison is such that the averaged path is in a direction towards the nearest vison. The annihilation rate in this regime is bounded by the terminal velocity of visons, which is one lattice spacing per MC~step. 

An exact analytical theory for such biased motion right after the early stage, however, is rather complicated. Nonetheless, the power-law decay of visons found in this regime can be understood from a mean-field type treatment~\cite{Hart2019}. Since the time taken to travel under the terminal velocity to a nearby vison is proportional to their distance, one can introduce a timescale $\tau_{\rm tv} \propto \ell $, where $\ell \sim 1/ \rho^{1/d}$ is the average distance between visons and $d$ is the spatial dimension. Treating this timescale as an effective lifetime in a rate equation, one obtains
\begin{align}
        \frac{d\rho}{dt} = -\frac{\rho}{\tau_{\text{tv}}} \propto - \rho^{1+ d^{-1}},
\end{align}
which can be readily integrated to yield an asymptotic behavior $\rho \sim t^{-d}$. In our case, this agrees well to the~$t^{-2}$ annihilation dynamics of visons observed in our kMC simulations in Fig.~\ref{fig:1}. When the vison density becomes very dilute, the finite-size effect of the long-range interaction comes to play (as shown in the inset of Fig~.\ref{fig:Vr} \cite{Note1}), reducing the strength of the interaction, so thermal fluctuations gradually overcome the ``terminal velocity" effect at the low density limit, and hence we partially recover the $~t^{-1}\ln t$ behavior toward the end of the annihilation process.  

    \begin{figure}[b]
    \includegraphics[width=\linewidth]{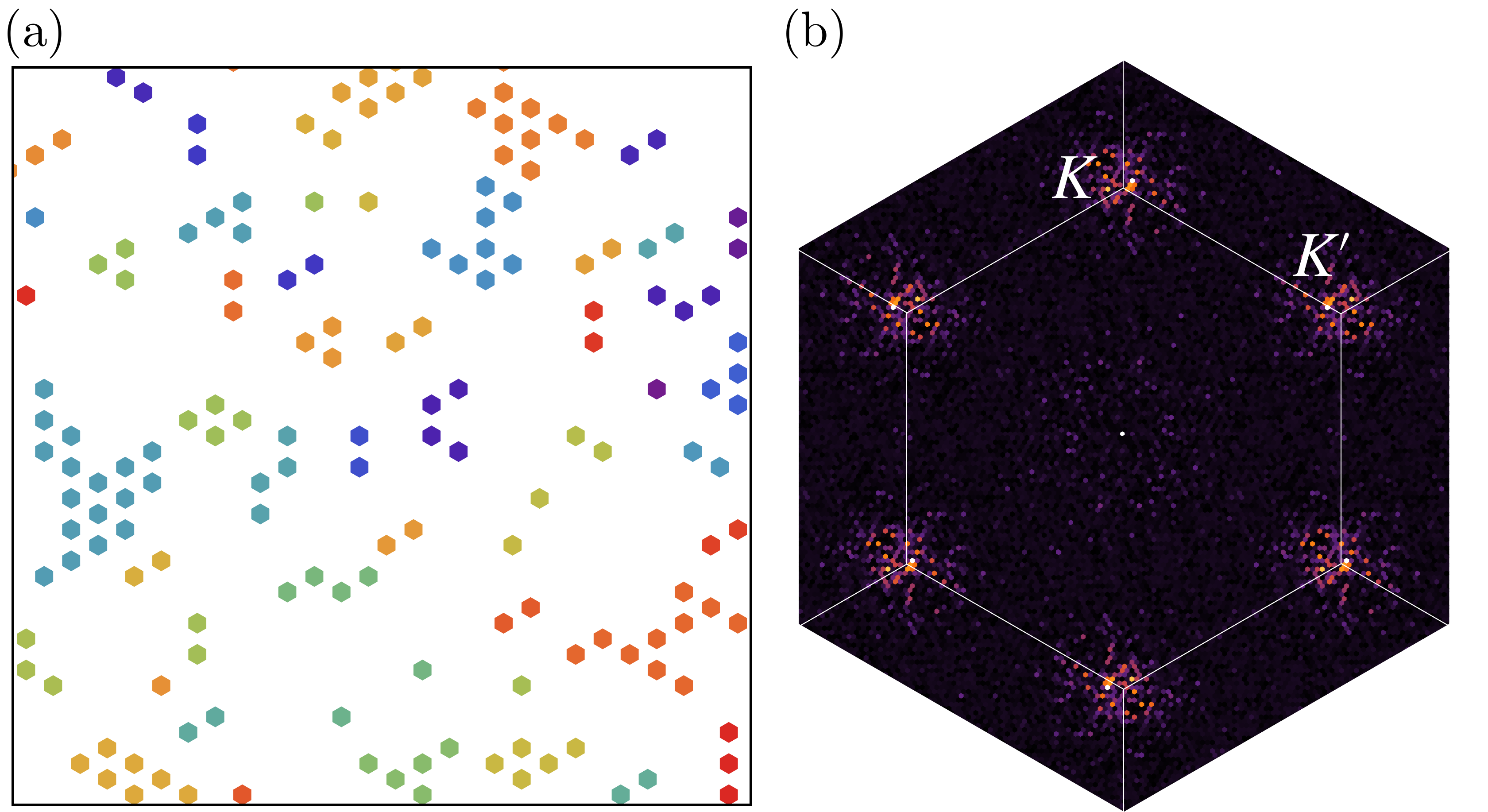}
        \caption{\label{fig:2} (a) Visons are frozen into small clusters of crystals (visons belonging to the same cluster are colored in the same color) at $T/|J|=10^{-3}$ (snapshot taken from an $N=90\times 90$ system at $t = 80$). (b) The static structure factor $S(\mathbf{q})$ of the entire $N=90\times90$ system: the bright intensities are centered around $K$ and $K'$ points on the boundary of the Brillouin zone reflecting the $\sqrt{3}\times\sqrt{3}$ structure of vison crystals.}
    \end{figure}

{\it Metastable vison crystals} -- At low quenching temperatures $T \lesssim T_{\rm freezing}$, the vison density quickly freezes to a nonzero value after the initial exponential decay; see the curve corresponding to $T/|J|\sim 10^{-3}$ in Fig.~\ref{fig:1}. This freezing behavior can be attributed to the trapping in local minima of the complex energy landscape of the multi-vison interactions. As discussed above, the pair potential exhibits oscillations with period as short as one lattice constant on top of the long-range attraction tail. In particular, two visons which are next-nearest-neighbor (NNN) to each other are more stable than a nearest-neighbor (NN) pair; their corresponding energy gain compared with two isolated visons are $\Delta E_{\rm NNN} \approx -0.079$ and $\Delta E_{\rm NN} \approx -0.044$~\cite{Kitaev2006}. This suggests an energy barrier of $\Delta V \approx 0.035$ that prevents a pair of visons from reaching each other. Our analysis shows that the probability of overcoming this barrier becomes less than 1\% when the post-quench temperature is below $T_{\rm freezing} \approx 0.006|J|$, which explains the freezing behavior \cite{Note1}.

    \begin{figure}[t]
        \centering
        \includegraphics[width=0.9\linewidth]{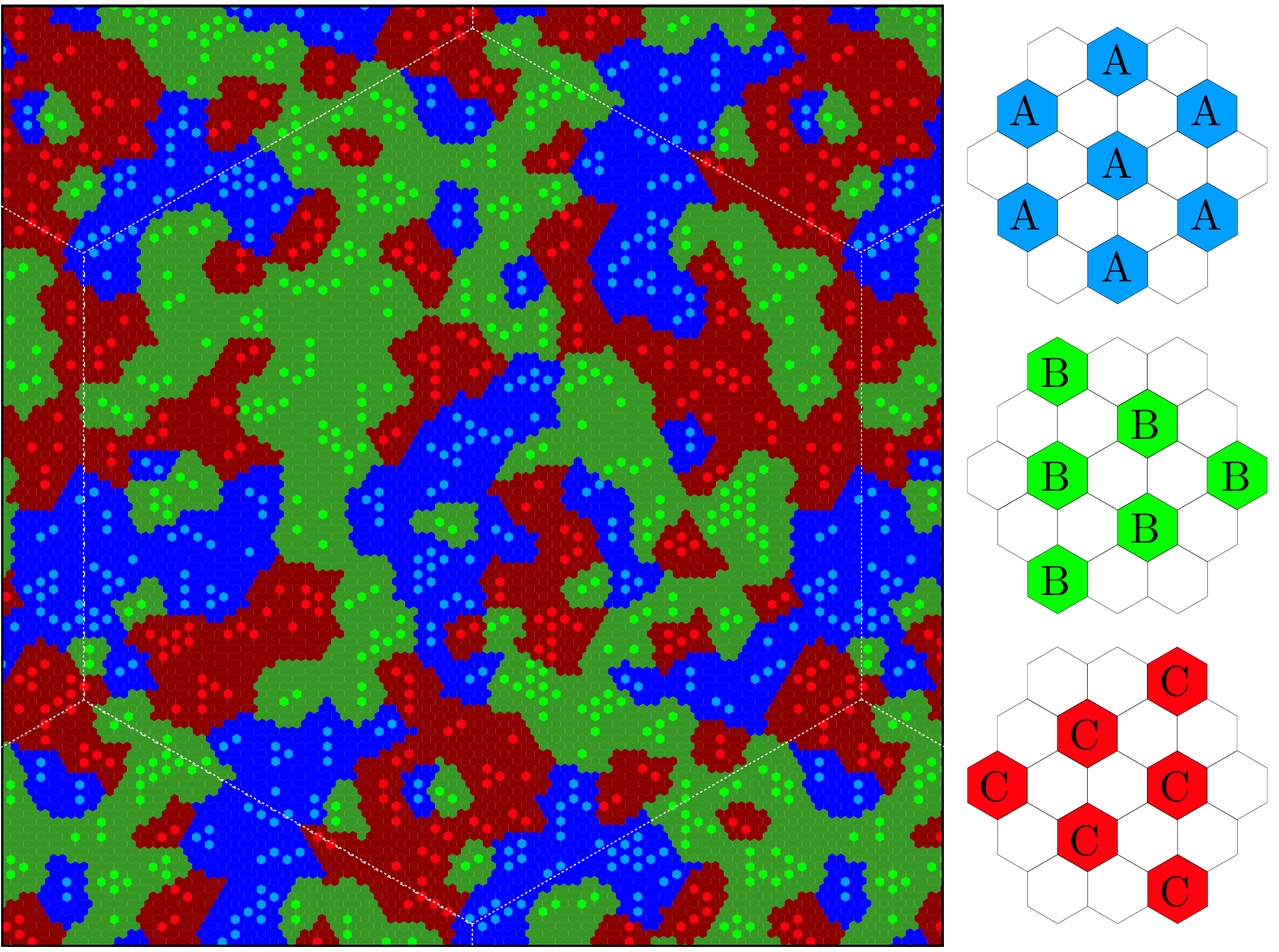}
        \caption{\label{fig:3} The system is partitioned into three domains (A, B, and C) based on the sublattice where the nearest vison resides (shown as bright hexagons in each domain, and three types of $\sqrt{3}\times\sqrt{3}$ vison crystals defined on the right). Vison crystals within the same domain form super-clusters.}
    \end{figure}

The local energy minimum associated with a pair of next-nearest-neighbor visons also suggests a metastable $\sqrt{3}\times \sqrt{3}$ pattern, which indeed corresponds to vison configurations that cost least energy. Detailed examinations reveal that the frozen states in our simulations are composed of fragmented vison crystals of a tripled unit cell as shown in Fig.~\ref{fig:2}(a). The emergence of the $\sqrt{3}\times \sqrt{3}$ vison crystals is also confirmed by the structure factor $S(\mathbf{q})$ of the frozen states which exhibits bright spots concentrated around $K$ and $K'$ corners of the Brillouin zone; see Fig.~\ref{fig:2}(b).

It is worth noting that since visons are energy costly in the Kitaev model, the $\sqrt{3}\times \sqrt{3}$ vison crystal is a metastable state caused by the arrested vison dynamics. This is in stark contrast to equilibrium vison crystals which are energetically stabilized by additional terms, e.g., four-spin interactions in modified Kitaev models~\cite{Zhang2019}. On the contrary, as mentioned above, the emergence of vison crystals here is attributed to the complex energy landscape of the multi-vison interactions through a self-sustained nonequilibrium process. 
To see this, we first notice that a $\sqrt{3}\times\sqrt{3}$ vison crystal breaks the $\mathbb{Z}_3$ sublattice symmetry of the triangular lattice, so each vison crystal can only reside on one of the three sublattices: A, B, or C. By introducing a three-state Potts variable $p = $ A, B, or C to every lattice site based on the sublattice of the closest vison crystal, the system can be partitioned into three types of domains, which can be viewed as superclusters of the vison crystals, as shown in Fig.~\ref{fig:3}. Our simulations found a rather slow growth of the $\mathbb{Z}_3$ domains, indicating a slow aggregation of vison crystals with their own species. A similar coarsening of super-clusters was observed in the relaxation dynamics of Falicov-Kimball model~\cite{Zhang2022}.

    \begin{figure}[b]
        \includegraphics[width=\linewidth]{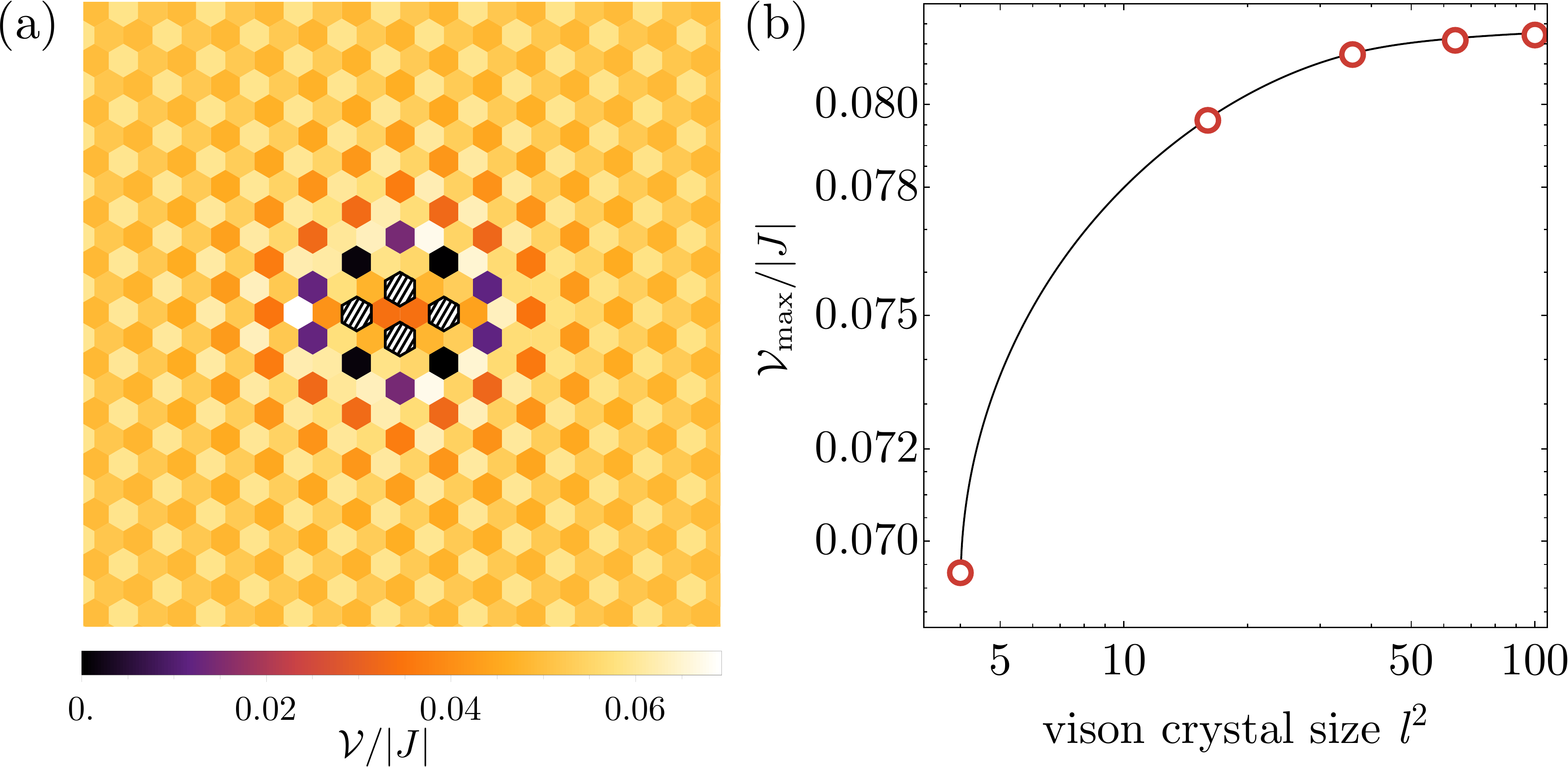}
        \caption{\label{fig:4} (a) Potential energy barrier $\mathcal{V}(x,y)$ created by a small $l^2=2\times 2$ vison crystal (shaded hexagons) embedded in a $N=120\times 120$ system. (b) The maximum potential energy barrier $\mathcal{V}_{\text{max}}$ versus the size of vison crystals.}
    \end{figure}
    
Interestingly, the inception of the vison crystals is also the source of the arrested dynamics of visons. This is because the presence of a vison crystal on one sublattice, even a small one, creates a concomitant $\sqrt{3}\times \sqrt{3}$ staggered potential field $\mathcal{V}(x, y)$ which favors the same sublattice in its neighborhood; see Fig.~\ref{fig:4}(a). This staggered potential, originated from the short-period oscillatory behavior of vison-vison interaction, thus creates energy barriers on the other two sublattices that suppress the diffusive motion of visons. The depth of the potential barriers is further found to be enhanced with increasing size of the central vison crystal, as shown in Fig.~\ref{fig:4}(b). While the reduced diffusivity prevents wandering visons to reach the central crystallite, thus limiting the size of vison crystals, the fact that they are trapped in the same sublattice nonetheless leads to the growth of the super-cluster and the coarsening of the~$\mathbb{Z}_3$ domains.

\begin{figure}[t]
        \centering
        \includegraphics[width=\linewidth]{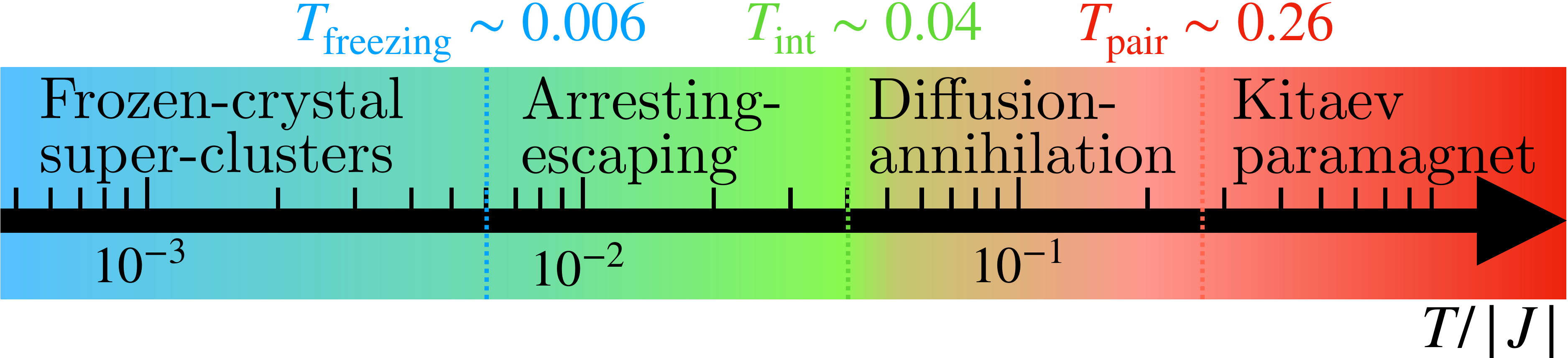}
        \caption{\label{fig:5} Different low-energy scales asscioated with the thermal quench dynamics of visons.}
    \end{figure}

{\it Discussion and outlook} --  
    The various low-temperature scales and corresponding thermal quench dynamics for visons in the gapless Kitaev spin liquid are summarized in Fig.~\ref{fig:5}. Above the pair-production temperature $T_{\rm pair} = 0.26|J|$, the detailed balance between thermal activation and pair-annihilation leads to an equilibrium state with finite vison density, also known as the ``Kitaev paramagnet" \cite{Nasu2014,Do2017}. As temperature is further lowered below $T_{\rm pair}$, spontaneous creation of visons is exponentially suppressed and the dynamics of residual visons can be described by the single-species $A + A \rightarrow \emptyset$ reaction-diffusion process, leading to a $\rho \sim t^{-1}\ln t$ decay of vison density. The attractive interactions between visons begin to take hold at temperatures below $T_{\rm int} \approx 0.04|J|$ and transform the diffusion-limited annihilation to a terminal-velocity limited one with a power-law decaying vison density $\rho \sim t^{-2}$. Finally, for quench temperature below $T_{\rm freezing} \approx 0.006|J|$, visons can hardly escape from trapping potential and gradually form super-clusters of metastable $\sqrt{3}\times\sqrt{3}$ vison crystals, which are purely of dynamic origin and do not require the introduction of extra couplings in the system.

    Our work reveals rich post-thermal-quench dynamics of visons in the gapless Kitaev spin liquid, arising from the interplay between thermal fluctuations and the inherent interactions of visons. One intriguing nonequilibrium phenomenon is the coarsening of the super-clusters of vison crystals. Since the tripled unit cell of $\sqrt{3}\times \sqrt{3}$ crystallization breaks a $\mathbb{Z}_3$ sublattice symmetry, the growth dynamics of super-clusters is expected to be similar to that of the 2D three-state Potts model. Our preliminary results, however, indicate a similar freezing behavior of the coarsening $\mathbb{Z}_3$ domains. A detailed study of the coarsening dynamics of super-clusters requires simulations of larger system sizes and longer timescales, which would be infeasible with ED-based kMC simulations. This computational difficulty can potentially be overcome by recent advances in machine learning (ML) force field methods that reduce computational complexity from $O(N^3)$ to $O(N)$ \cite{Behler2007,Bartok2010,Zhang2018,Suwa2019}.

    The ML methods will also be important for studying thermal quenches of 3D Kitaev spin liquid in, e.g.,  the hyperhoneycomb lattice~\cite{Lee14,Kimchi14}, where the enlarged dimensions require more computational overhead. Moreover, contrary to the 2D case studied here, the spin liquid phases in the 3D Kitaev systems at low temperatures are always distinguished from the high-temperature paramagnet by a phase transition~\cite{Nasu2014}. As there is no obvious symmetry breaking involved, the exact nature of the phase transitions requires further investigations. From the dynamical point of view, the 3D Kitaev system might exhibit qualitatively different quench dynamics due to the presence of a true phase transition.

    Finally, the quench dynamics of visons or, more generally, anyon-like defects, in a gapped Kitaev spin liquid remains an open question. While thermal coupling typically destroys the quantum coherence, justifying an effective kMC description, the non-abelian nature of these defects could lead to distinct non-classical late-time dynamics. This is because some of the quantum information about the state is topologically protected from the environment. For example, in addition to diffusion and pair-annihilation, these topological defects also exhibit nontrivial species transmutation governed by well-defined fusion rules. Previous studies so far have focused on 1D systems, partly because of the computational difficulty~\cite{Nahum2020,Lin2021}. Further studies of non-abelian defects in high-dimensional spin liquids, either based on empirical models or Kitaev spin liquids with ML-enabled large-scale simulations would open a new chapter in the field of phase ordering dynamics.

{\it Acknowledgment} --
    The authors would like to thank Shang-Shun Zhang  for discussions and Natalia B. Perkins for her proofreading of the manuscript and valuable suggestions. This work is supported by the US Department of Energy Basic Energy Sciences under Award No. DE-SC0020330. The authors also acknowledge the support of Research Computing at the University of Virginia. 
    
\bibliography{main.bbl}

\end{document}


\title{Supplemental Materials for \protect\\``Thermal Quench Dynamics of Visons in Gapless Kitaev Spin Liquid"}
    \author{Yang Yang}
    \affiliation{Department of Physics, University of Virginia, Charlottesville, Virginia, 22904, USA}
    \author{Gia-Wei Chern}
    \affiliation{Department of Physics, University of Virginia, Charlottesville, Virginia, 22904, USA}
    \maketitle

    \setcounter{figure}{0}
    \setcounter{equation}{0}
    \renewcommand{\theequation}{S\arabic{equation}}
    \renewcommand{\thefigure}{S\arabic{figure}}
    
    \section{S1. Details on the simulation method}
        \subsection{A. Diagonalization of Kitaev Hamiltonian at finite temperature}
        In this section, we present details of our simulation method for the post-thermal-quench dynamics of visons in the Kitaev model. 
        We start with the Kitaev model in the isotropic limit \cite{Kitaev2006}
        \begin{align}\label{eqn:Kitaev}
            \mathcal{H}_{\text{K}}=-J\sum_{\alpha=x,y,z}\sum_{\langle jk\rangle_\alpha}\sigma^\alpha_j\sigma^\alpha_k,
        \end{align}
        where $\sigma_j^x,\sigma_j^y,\sigma_j^z$ denote Pauli matrices representing the original spin-1/2 degrees of freedom in the Kitaev model. 
        Using Majorana representations of Pauli matrices we can rewrite Eq.~(\ref{eqn:Kitaev}) into the quadratic form
            \begin{align}\label{eqn:majorana}
            \tilde{\mathcal{H}}_{\text{K}}=i J\sum_{\langle jk\rangle}u_{jk}c_j c_k,
        \end{align}
        where $c_jc_k$ describes the hopping terms of the matter Majorana fermions and $u_{jk}$ is a $\mathbb{Z}_2$ gauge field given by the expectation value of the gauge Majorana fermions $i\langle b_j^\alpha b_k^\alpha\rangle$ which takes value $\pm 1$ ($u_{jk}$ is defined to connected the even sublattice $j$ site to the odd sublattice site $k$ to avoid the sign ambiguity). The exact solution of the Kitaev model amounts to Eq.~(\ref{eqn:majorana}) being a free fermion Hamiltonian for the matter Majorana fermions. 
        Then the product of the $\mathbb{Z}_2$ gauge field around each hexagon loop defines a gauge-invariant $\mathbb{Z}_2$ flux degree of freedom 
        \begin{align}
            W_p=\prod_{\langle jk\rangle\in\hexagon}u_{jk},
        \end{align}
        which also takes values $\pm 1$.
        Since it is gauge-invariant, once the flux configuration of the system is given, the spectrum of Eq.~(\ref{eqn:majorana}) is determined.  
        Reciprocally, we can obtain the energy of a flux configuration by computing Majorana fermion's spectrum under just one of many gauge-equivalent gauge fields.
        
\begin{figure}[ht]
            \includegraphics[width=\linewidth]{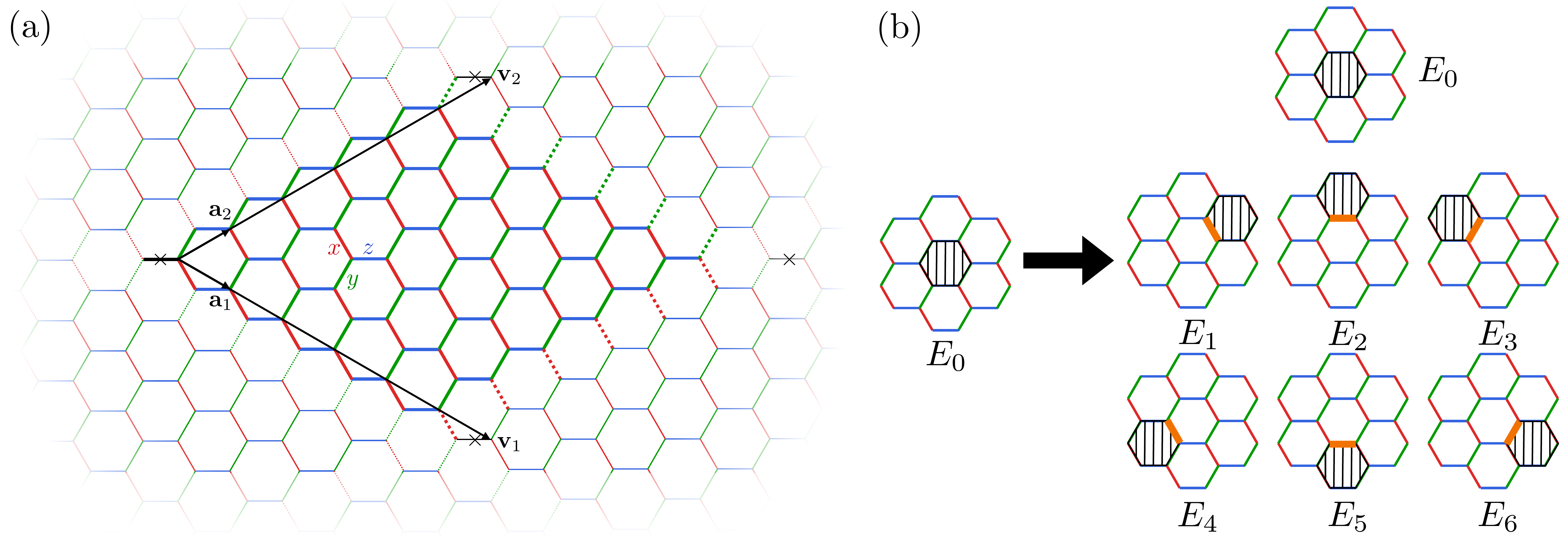}
            \caption{\label{fig:S1} (a) An example of a $N=6\times 6$ cluster with the periodic boundary condition generated by the spanning vectors $\mathbf{v}_1$ and $\mathbf{v}_2$. Solid bold bonds are the bonds in the $N=6\times 6$ cluster. Dashed bonds are the periodic boundary bonds connecting The broken bond is denoted by ``$\times$". The simulation is performed on the $N=32\times 32$, $N=60\times 60$, and $N=90\times 90$ clusters with the same setup. (b) Seven outcomes of a MC update for a single vison originally sitting in the middle determined with the probability given by Eq.~(\ref{eqn:MC_update}); the hopping of the vison is achieved by flipping $u_{ij}$ on the orange bonds.}
        \end{figure}

        Next, we take the advantage of the sublattice symmetry of the honeycomb model by rewriting Eq.~(\ref{eqn:majorana}) in terms of a $2L\times 2L$ matrix $M$ which takes a block off-diagonal form
        \begin{align}
            M=\left(\begin{array}{c|c}
                0 & A\\
                 \hline
                A^\dagger & 0
             \end{array}\right),
        \end{align}
        Note that characteristic equations for eigenvalues have $\det(M-\lambda\mathbb{1})=\det(A^\dagger A-\lambda^2\mathbb{1})$.  
        The eigenvalues of $M$ are given by the square root of eigenvalues of $A^\dagger A$. 
        Hence, the sublattices in the non-Bravais honeycomb lattice do not increase the computational complexity, i.e.,
        diagonalization of a Hamiltonian on a $N=L\times L$ non-Bravais takes same amount of time as diagonalization of a Hamiltonian on a Bravais lattice. 
        Once we fix a set of gauge variables $u_{jk}$ with a given vison configuration, we can diagonalize Hamiltonian in terms of the complex fermion operators $a_k^\dagger$ and $a_k$,
        \begin{align}\label{eqn:diagoalized}            \tilde{\mathcal{H}}_{\text{diag}}=\sum_{k}^{N}\epsilon_k\left(a_k^\dagger a_k-\frac{1}{2}\right), 
        \end{align}
        where $\epsilon_k$ are the positive eigenvalues of $M$. 
        By replacing $a_k^\dagger a_k$ with a Fermi-Dirac distribution function $1/(e^{ \epsilon_k/T}+1)$ (i.e., assuming Majorana fermions relax much faster than visons), we obtain the thermal expectation values of a vison configuration at the corresponding temperature $T$
        \begin{align}\label{eqn:vison_config}
            E_{\text{visons}}(T)=-\sum_{k}^{N}\frac{\epsilon_k}{2}\tanh\left(\frac{\epsilon_k}{2T}\right).
        \end{align}
        Extra caution must be exercised when computing the energy of a given vison configuration. The Majorana representation of Pauli matrices enlarges the Hilbert space of the original spin Hamiltonian, so not all states in Eq.~(\ref{eqn:diagoalized}) correspond to the actual eigenstates in the Eq.~(\ref{eqn:Kitaev}). This redundancy of unphysical states in the fermion Hamiltonian can be eliminated by imposing a parity condition on the state \cite{Pedrocchi2011}. The parity condition ensures that only physically relevant states are considered when determining the energy spectrum of the system. A more practical way of implementation of this parity requirement is to remove a bond from the system such that a Majorana zero mode can be obtained on this dangling bond to satisfy the parity requirement of the whole wave function \cite{Zschocke2015,Zhang2019,Feng2020}. An example of such implementation on a small system is shown in Fig.~\ref{fig:S1} (a). We compare the ground state energy difference between the system with a broken bond and without the broken bond, which is at the order of $\sim 10^{-4}|J|$ for the system of size $N=60\times 60$. Hence such treatment will not affect the results of our work.
                
        \subsection{B. Quantum mapping for open-system adiabatic dynamics }
        The thermal hopping dynamics of visons in an open environment can be described by quantum dynamical maps by Kraus operators \cite{Kraus1983}. The most general dynamical map for the density matrix of the entire system
        \begin{align}
            \rho_K(t+\delta t)=\sum_\eta \mathcal{K}_\eta\rho_K(t)\mathcal{K}_\eta^\dagger,
        \end{align}
        where $\{\mathcal{K}_\eta\}$ forms a set of Kraus operators satisfying $\sum_\eta \mathcal{K}_\eta^\dagger\mathcal{K}_\eta=\mathbb{1}$, maps the initial density matrix $\rho_K$ to $\rho_K'$ after some characteristic time $\delta t$. Under adiabatic approximation, we partition the full density matrix of our system as
        \begin{align}
            \rho_{K} = \rho_{\text{V}}\otimes\rho_{\text{M}},
        \end{align}
        where $\rho_{\text{V}}$ gives the density matrix of the slow vison degrees of freedom, and $\rho_{\text{M}}$ gives the density matrix of the fast Majorana degrees of freedom. We first consider the fast dynamics of Majorana degrees of freedom are described by the Kraus mapping $\{\mathcal{K}_{\mathrm{fast},M}\}$ with characteristic time $\tau_M$. Application of $\{\mathcal{K}_{\mathrm{fast},M}\}$ quickly brings Majorana degrees of freedom in thermal equilibrium with the external environment. For a Majorana Hamiltonian with $2N$ degrees of freedom, this can be done through $N!+1$ channels of Kraus operators in a similar manner to thermalizing an open two-level system \cite{Cherian2019}. For the slow vison dynamics, we consider the Kraus mapping $\{\mathcal{K}_{\mathrm{slow},V}\}$ with charateristic time $\tau_1\gg \tau_M$,  where each $\mathcal{K}_{\mathrm{slow},V}$ takes the form
        \begin{align}\label{eqn:Kvison}
            \mathcal{K}_{\mathrm{slow},V} = c(\mathcal{C}', \mathcal{C})|\mathcal{C}'\rangle\langle \mathcal{C}|,
        \end{align}
        where vison configuration $\mathcal{C}'$ and $\mathcal{C}$ are differed by only single vison hopping, and transition amplitude $c(\mathcal{C}', \mathcal{C})$ depends on the underlying Majorana spectra of these two vison configurations. Application of $\{\mathcal{K}_{\mathrm{slow},V}\}$ to the vison degrees of freedom assumes tracing out the Majorana degrees of freedom 
        \begin{align}
            \rho_{\text{V}}(t)= \mathrm{Tr}_{M}\rho_K=\sum_\mathcal{C}\pi(\mathcal{C};t)|\mathcal{C}\rangle\langle \mathcal{C}|,
        \end{align}
        by treating thermalized Majorana degrees of freedom like external heat bath, and $\pi(\mathcal{C};t)$ corresponding to the probability distribution for vison configurations. Notably, adiabtic approximation also assumes $\rho_{\text{V}}$ is fully diagonal during entire process, as environmental interactions quickly destroy vison superposition coherence. This is justified since the decoherence time $\tau_2$ is typically orders of magnitude smaller than the dynamical timescale $\tau_1$, analogous to two distinct relaxation times in Bloch spin dynamics \cite{Bloch1946}. Therefore, the adiabatic dynamics of visons under Kraus operators becomes the time evolution of probability distribution $\pi(\mathcal{C};t)$. Since the system will eventually reach thermal equilibrium in an open environment, $\pi(\mathcal{C};t)$ must evolve into Gibbs distribution,
        \begin{eqnarray}
        \pi^{\mathrm{th}}(\mathcal{C})\propto e^{-\frac{E(\mathcal{C},T)}{T}}. 
        \end{eqnarray}
        When the characteristic time scale arising from the coupling to the thermal environment satisfies $\tau_1\gg \tau_2 \gg \tau_M$, we only need to focus on the relaxation dynamics of visons, and model the application of the Kraus operators in Eq.~(\ref{eqn:Kvison}) as a Markov chain process of changing vison configuration density 
        \begin{eqnarray}
    \pi(\mathcal{C}, t+\delta t) = \sum_{\mathcal{C}'} P(\mathcal{C}' \to \mathcal{C}) \pi(\mathcal{C}'; t),
        \end{eqnarray}
        where $\delta t$ is of the order of $\tau_1$. This allows us to study the dynamical relaxation of visons through kinetic Monte Carlo (kMC) simulations.
        
        Assuming the interaction between the system and the environment is local, we construct our kMC simulations by sampling nearest-neighbor hopping events of visons. This is done by considering all the outcomes of a single vison making nearest-neighbor hopping shown in Fig.~\ref{fig:S1} (b). The hopping of visons is achieved by flipping the $\mathbb{Z}_2$ gauge field $u_{ij}$ on the bond shared by two hexagons. Each outcome associated with a energy of the system computed by Eq.~(\ref{eqn:vison_config}), and we label them as $E_0$, $E_1$, $\dots$, $E_6$. If the neighboring site is already occupied with another vison, then we compute the energy of the system after the hopping as they have annihilated. Whether a vison decides to stay at original place or hop to its nearest neighbors depends on the Boltzmann factor, $\exp(-E_k/T)$, so the probability of each outcome associated with energy $E_k$ is determined by
        \begin{align}\label{eqn:MC_update}
            P(E_k)=\frac{\exp\left(-E_k/T\right)}{\sum_{j=0}^6\exp\left(-E_j/T\right)}.
        \end{align}
        This vison hopping probability is directly analogous to the spin-flipping probability in the Glauber dynamics for simulating kinetic Ising model \cite{Glauber1963}. Thus, we obtain a kMC simulation scheme for vison relaxation dynamics.

        However, an update of a vison's location still requires comparing the energy of the current vison configurations to the energies of vison configurations after all possible hoppings, necessitating six full diagonalizations of the Majorana fermion Hamiltonians. This computationally expensive process, especially for large system sizes, can be significantly accelerated using GPU parallelization \cite{Besard2018}. Therefore, we can simulate the system with sizes that would typically require a supercomputer cluster, achieving equivalent results within a similar timeframe on a single GPU.
    
        We prepare the system at the infinitely high temperature by randomly sampling the gauge field variables $u_{jk}$ such that we obtain a random flux configuration with vison density roughly at $\rho=0.5$ to start the simulation. The simulations are ran on the system of the size $N=32\times 32$, $N=60\times 60$, and $N=90\times 90$, respectively. 
    \section{S2. Examination of finite-size effects}
\begin{figure}[ht]
            \includegraphics[width=\linewidth]{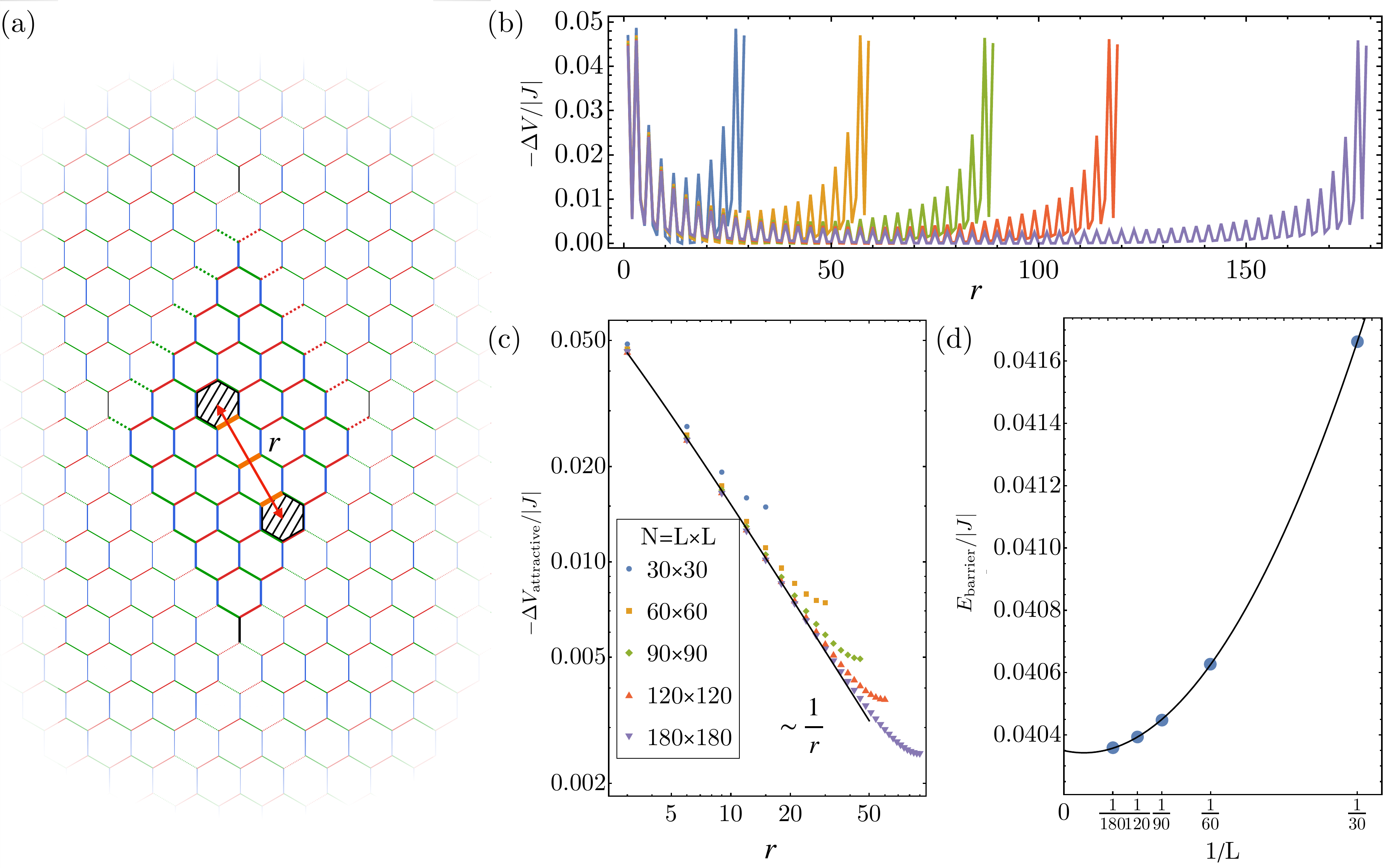}
            \caption{\label{fig:S2} (a) Configuration for measurement of the two-vison interaction on the system with periodic boundary condition: two visons separated with the distance $r$ are created by flipping all $\mathbb{Z}_2$ gauge variables $u_{ij}$ on orange bonds between them. (b) Two-vison interaction strength $V$ with respect to their separation $r$ with the zero point energy fixed at $r =L/2$, and the corresponding colors of different system sizes are labeled in (c).  (c) The $1/r$ dependence of the long-range attractive interactions between two visons. (d) Extrapolate the potential barrier with the system size $1/L$; the fitting is done with $a+b/L+c/L^2$ giving $a=0.0403(5)$. }
        \end{figure}

       In this section, we examine possible finite-size effects in our simulations. We demonstrate that our main results, presented in Fig.~1 of the main text for the $N=60\times 60$ system, are not qualitatively affected by these finite-size effects. First, we compare two-vison interactions $\Delta V$ on various system sizes by computing the energy of different two-vison configurations with a two-vison separation of $r$ [Fig. \ref{fig:S2} (a)]. From Fig.~\ref{fig:S2} (b), we can see that the interaction between two visons is oscillatory within the short range, but is overall attractive across the long distance. The depth of the potential ($0.04|J|$, corresponding to the binding energy of two visons) is nearly identical across different system sizes. We can fit the long-range attractive interaction with $1/r$ [Fig.~\ref{fig:S2} (c)] by taking peaks of Fig. \ref{fig:S2} (b) up to half of the system size, since the system is wrapped around with the periodic boundary condition. The long-range interaction fits well with the $1/r$ dependence except at the tail of the fitting, where the the finite size effect comes to play. This explains the recurrence of the $t^{-1}\ln t$ behavior from the anomalous relaxation $t^{-2}$ when the density of visons becomes low at the intermediate temperature regimes. Similarly, we can also look at the finite size effects for the short-range potential barrier $E_{\rm barrier}\equiv \Delta V(r=2) - \Delta V(r=3)$. The results are shown in \ref{fig:S2} (d) with the extrapolation $1/L\rightarrow 0$ done using $a+b/L+c/L^2$. The extrapolation gives $a=0.0403(5)$, which is of the order $10^{-4}|J|$ again different from the interaction for the system $N=60\times 60$. Therefore, our results should not be affected by these finite-size effects.
        
\begin{figure}[ht]
            \includegraphics[width=0.5\linewidth]{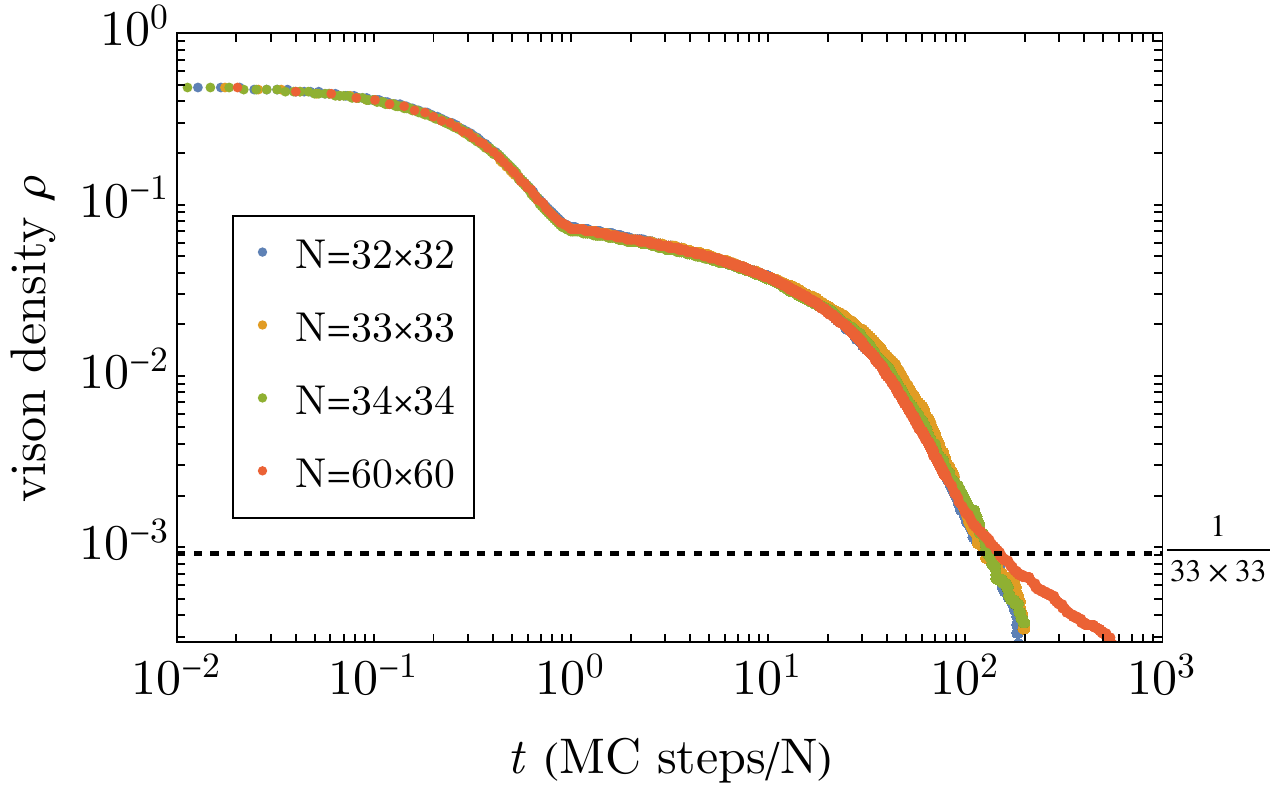}
            \caption{\label{fig:S3} Comparison of the anomalous relaxation dynamics of visons in the systems with the sizes $N=32\times 32$, $N=33\times 33$, $N=34\times 34$, and $N=60\times 60$ at $T/|J|=10^{-2}$. All show $t^{-2}$ anomalous relaxation dynamics. The finite-size effects appear when $\rho$ is below $1/N$.}
        \end{figure}

        We also check the diffusion-annihilation dynamics for visons on various system sizes at $T/|J|=10^{-2}$. Since almost all visons are annihilated around $t=100$, we see finite-size effects arise around $t=100$ where $\rho<1/N$ for the smaller systems ($N=32\times 32$, $N=33\times 33$, and $N=34\times 34$). It is worth noting that if the linear dimension of the system is not a multiple of $3$, the discrete spectrum of the Hamiltonian will not include the Dirac points ($K$ and $K'$ points in the Brillouin zone) of the continuous spectrum in the thermodynamic limit. However, Fig.~\ref{fig:S3} shows the anomalous diffusion-annihilation is present in all systems with linear dimension $3L$, $3L+1$ and $3L+2$, and displays nearly the same anomalous behaviors as the behavior in a much larger system $N=60\times 60$. Therefore, we can conclude that the finite-size effects are not affecting results in our main text.
        
    \section{S3. More low-temperature results}
\begin{figure}[ht]
            \includegraphics[width=\linewidth]{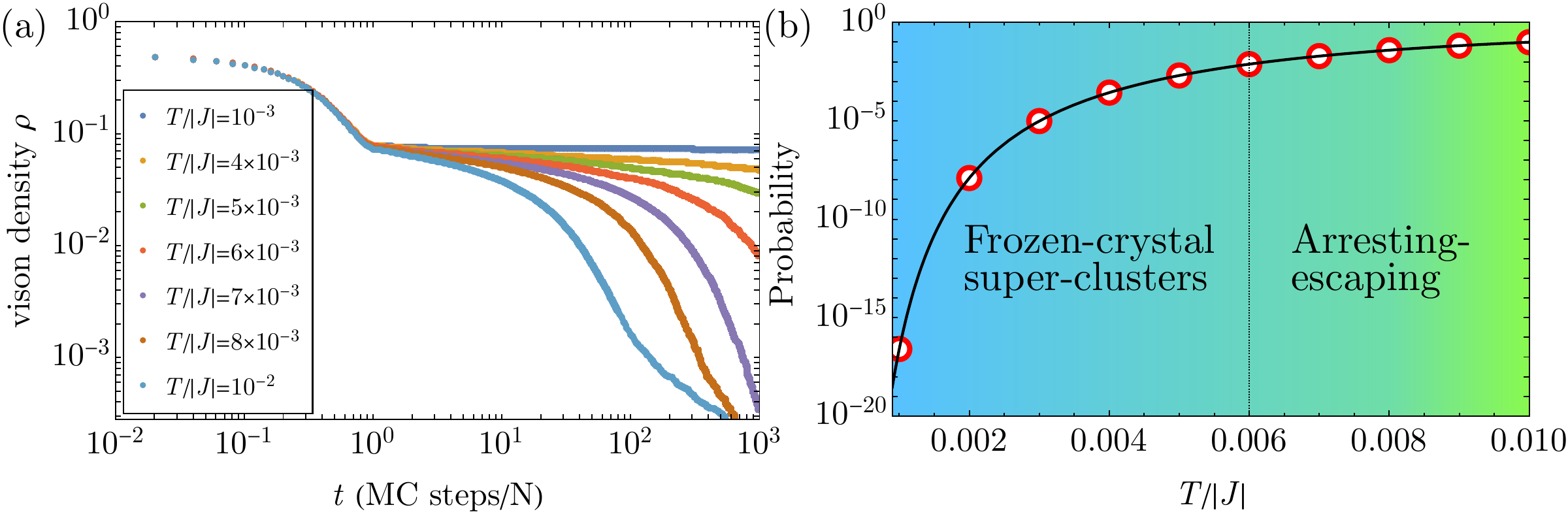}
            \caption{\label{fig:S4} (a) The time evolution of vison density at low temperatures: the temperature dependence shows a crossover behavior between the metastable vison crystals and anomalous relaxation dynamics. (b) The probability for a vison to overcome a potential barrier of approximately 0.04 $|J|$ and hop to a neighboring site is shown as a function of temperature. The crossover temperature, $T/|J|\sim 0.006$, is marked by the point at which the probability for a trapped vison to hop reaches about $1\%$. }
        \end{figure}

        Last, we provide more numerical results for post-thermal-quench dynamics for visons at the low-temperature regimes. Fig.~\ref{fig:S4} (a) shows a crossover behavior between the metastable vison crystals and anomalous relaxation dynamics, when $T/|J|$ is increased from $10^{-3}$ to $10^{-2}$. Hence there is no sharp transition temperture that marks the boundary between two dynamics of the system. However, we can estimate a crossover temperature from the probability for a single vison to hop over a potential barrier of $0.04|J|$, which is the maximum repulsive interaction strength between two visons obtained from the finite-size extrapolation. This probability is estimated by
        \begin{align}
            P_{\mathrm{hop}} \sim \frac{6\exp(-\Delta V/T)}{6\exp(-\Delta V/T)+1},
        \end{align}
        according to Eq.~(\ref{eqn:MC_update}). The results are shown in Fig.~\ref{fig:S4} (b). Therefore, we can mark the crossover temperature of the system at $T/|J|\sim 0.006$ when the probability of the hopping is roughly $1\%$.
    \bibliography{supp.bbl}